# Time Division Multiplexing Ising Computer Using Single Tunable True Random Number Generator Based on Spin Torque Nano-Oscillator

Bolin Zhang[1,2], Yu Liu[1,2], Tianqi Gao[1,2], Deming Zhang[1], Weisheng Zhao[1,2] and Lang Zeng[1,2]
[1]Fert Beijing Institute, MIIT Key Laboratory of Spintronics, School of Integrated Circuit Science and Engineering, Beihang University, Beijing 100191, China
[2]Beihang Hangzhou Innovation Institute Yuhang, Xixi Octagon City, Yuhang District, Hangzhou 310023, China
Email: zenglang@buaa.edu.cn, weisheng.zhao@buaa.edu.cn

*Abstract*—Ising computer is a powerful computation scheme to deal with NP-hard optimization problems that cannot be efficiently addressed by conventional computers. A robust probabilistic bit (P-Bit) which is realized by a hardware entity fluctuating in time between -1 and 1 plays a key role in the success of Ising computer. Spintronics technology, such as stochastic nanomagnet, is recently proposed as a good platform for the hardware emulation of P-Bit. Here, we report, for the first time, a Time Division Multiplexing (TDM) Ising computer using single tunable true random number generator which is comprised of a Spin Torque Nano-Oscillator (STNO). First, the intrinsic frequency fluctuation of the STNO is utilized to design a simple digital true random number generator (TRNG). The true random number generator is further evolved into a tunable random number generator to act as a P-Bit. Second, in order to accomplish combinational optimization with our proposed P-Bit under Ising model, a novel incremental coupling rule is proposed. With such coupling rule between P-Bit array, high fidelity NOT and XOR logic gate is demonstrated. Third, it is proposed that our digital TRNG can be simply reused acting as a P-Bit array by time division multiplexing. The whole Ising computer can be implemented by one single STNO, and integer factorization of as high as 87% accuracy rate is achieved.

## I. Introduction

Recently, it is experimentally demonstrated that stochastic magnetic tunnel junctions (MTJ) can be used as a hardware entity of probabilistic bit (P-Bit) to comprise an Ising computer [1]. This maybe posted as the most advanced progress for physical implementation of Ising computer. However, there still exists three major drawbacks for stochastic MTJ based P-Bit device and array. 1) The stochastic MTJ based P-Bit needs a careful design of reading circuit [2]. The reading circuit should manage with the low Tunneling Magneto Resistance (TMR) ratio while it must not affect the vulnerable switching process of stochastic MTJ. 2) The stochastic MTJ is difficult to fabricate since it asks for very small magnetic anisotropy to keep energy barrier $\Delta E$ comparable with thermal energy. Tiny variation in energy barrier $\Delta E$ strongly affects the stochastic switching characteristic of MTJ. For this reason, the stochastic MTJ P-Bits used for Ising computation have to be calibrated one by one to get correct results. This calibration is not practical for large array. 3) For an Ising computer comprised of large P-Bit array, the coupling between each P-Bit poses another enormous challenge. It is known that $O(N^2)$ coupling connections are required for a P-Bit array with size of $O(N)$. Such squarely increasing coupling connections are unprocurable for large P-Bit array. In this work, we proposed a digital tunable TRNG based on STNO to act as a better P-Bit. The limitations of calibration and coupling connections are solved by time division multiplexing.

## II. Tunable TRNG Based on STNO

The structure of the STNO used in our work is shown in Fig. 1 [3]. It is basically the CoFeB/MgO/CoFeB sandwich structure which is almost the same as an ordinary MTJ devices. As shown in Fig. 1, a stable magnetic rotation can be sustained if the anti-damping Spin Transfer Torque term induced by injected spin polarized current and the Gilbert damping term can cancel each other. This stable magnetic rotation will cause a resistance variation since TMR effect. In our STNO, the fixed layer is made of in-plane magnetization and the free layer is with Perpendicular Magnetic Anisotropy (PMA). Such structure can provide the largest resistance variation leading to large RF output voltage [3]. The STNO device parameters are listed in Table I, and the oscillation frequency dependence on injected current is shown in Fig. 2.

The thermal noise will affect the stable magnetic rotation in STNO since the device dimension is only $\pi \times 50 \times 50 \times 2$ nm$^3$. The FFT analysis of the simulation results including thermal fluctuation is shown in Fig. 3. It is shown that the oscillation frequency spreads over a wide range. It implies that the oscillation frequency can be used as entropy source to generate random numbers. As shown in Fig. 4, the number of oscillations of our STNO with 1.6 mA injected current in every 100 ns is statistically collected. The occurrence probability for number of oscillations can be approximated by a Gaussian distribution with mean value around 466. The RF output voltage of the STNO is extracted and plotted in Fig. 5. The output voltage oscillates between 0.31 and 0.48 V.

Summarizing the above observations, we proposed our simple digital TRNG based on the STNO shown in Fig. 6. The oscillating output voltage is measured by a Schmitt trigger and converted into rail-to-rail full swing square wave triggering a CMOS counter by adding one. At the end of every period of 100 ns, the counter will generate a logic '-1' if its value is larger than a predefined threshold and reset its value to zero. Otherwise, it will generate logic '1' and then reset to zero. The simulation result of this digital TRNG is shown in Fig. 7. For a 50/50 percentage of '1' and '-1', the threshold is set to be 466. The generated random number sequence is fed into NIST test suite, and all the 11 NIST tests are passed as shown in Fig. 8. The threshold can be conveniently modified to change the occurrence probability of '1' and '-1'. The tunable occurrence probability of '1' is shown in Fig. 9, and it can be fitted well by the Sigmoidal function.

## III. Incremental Coupling Rule for Proposed P-Bit

With the proposed tunable TRNG as a P-Bit, it is ready to design an Ising computer. The electrically connected P-Bits forming a functional asynchronous Ising network is proposed in Ref. [1], and its implementation for the P-Bit in our work is shown in Fig. 10. The Hamiltonian of Ising Model for NOT logic gate is [4]

$$H_{NOT} = 1 - (-C_1 C_2)$$

Analogue to stochastic nanomagnet, the thresholds at time $t$ of the two P-Bits comprising of the NOT logic gate can be calculated as [4]

$$T_1(t) = T_0 - C_2(t - \Delta t) \times \Delta T$$
$$T_2(t) = T_0 - C_1(t - \Delta t) \times \Delta T \quad (1)$$

where $T_1/T_2$ is the threshold of P-Bit 1/2 at time $t$, $T_0$ is the threshold value of 466 which corresponds to 50% probability output, $C_1/C_2$ is the random number generated by P-Bit 1/2 at time $t-\Delta t$, $\Delta t$ is 100 ns, and $\Delta T$ is the coupling strength between the two P-Bit. According to Fig. 9, $\Delta T$ can be set as about 20.

However, contrary to stochastic nanomagnet, such coupling rule fails to give correct results and the accuracy rate is only 49% which means the results are almost purely stochastic. The reason is analyzed in details as shown by Fig. 11. The red region in Fig. 11 highlights the oscillatory wrong states. It can be understood as follows: When both of the two P-Bits are in state '1', the mutual coupling between them as shown by Fig. 10(b) will forcibly push them into state '-1'. For stochastic nanomagnet, one of the two P-Bits based on MTJ will arrive state '-1' before the other one. Once the faster P-Bit becomes state '-1', it will push the other one P-Bit back to state '1'. In this way, the Ising computer can jump out from the wrong '11' state and become the correct '-11' or '1-1' state. Since our proposed is digital tunable TRNG which generates random number in every 100 ns, the situation is totally different from that of P-Bits based on MTJ. When the Ising computer is in wrong '11' state, it will become still wrong '-1-1' state after 100 ns. Then it oscillates between wrong '11' and '-1-1' state. Since $\Delta T$ is set as about 20, it has little chance to jump out from this oscillatory wrong state. We must design a new coupling rule which works for our digital P-Bits.

As shown by the black line in Fig. 12, the old coupling rule will make the threshold jumps between $T_{max}$ and $T_{min}$. We observed that for threshold around $T_0$, the probability is around 50%. Thus, the probability of jumping out from the wrong state can be 50% if the threshold is set around $T_0$. Inspired by this observation, we designed a novel incremental coupling rule indicated by the blue line in Fig. 12. This new rule can be written as

$$T_1(t) = T_1(t - \Delta t) - C_2(t - \Delta t) \times \Delta T$$
$$T_2(t) = T_2(t - \Delta t) - C_1(t - \Delta t) \times \Delta T \quad (2)$$

Instead of being determined solely by the P-Bits output at time $t-\Delta t$ in the old rule, the threshold is incrementally changed from its previous value by step of $\Delta T$. The new rule introduces two additional parameters, the range between $T_{max}$ and $T_{min}$ and the step $\Delta T$, which can be tuned carefully for higher accuracy rate. For example, the accuracy rate is 74% when $T_{max} - T_{min} = 21$ and $\Delta T = 7$ as shown in Fig. 13. This accuracy rate is much higher than 67% got from MTJ based P-Bit calculation. In Fig. 14, the accuracy rate of 64% is demonstrated for XOR logic gate with our incremental coupling rule.

### IV. Time Division Multiplexing Ising Computer

A significant feature of our Ising computer is its asynchrony. Utilizing this feature, a time division multiplexing Ising computer can be designed as shown in Fig. 15. The TDM workflow is sketched in Fig. 16. The single P-Bit based on tunable TRNG first computes as P-Bit 1. After 100 ns, its random number is generated and stored in memory. Then the Threshold Setting signal in Fig. 6 is enabled, and the threshold for P-Bit 2 is reloaded from the memory into the CMOS counter. Now the P-Bit device computes as P-Bit 2. After 100 ns, its random number is generated and stored in memory too. Then the threshold of P-Bit 1 is read from the memory, calculated by the incremental coupling rule and reloaded into the CMOS counter. The P-Bit device computes as P-Bit 1 again. With this TDM workflow, all of the calculation can be done by one single P-Bit device and no physical connections between P-Bit array needed any more.

The simulation result for a NOT logic gate by this TDM workflow is shown in Fig. 17. The Yellow/Blue region is for P-Bit 1/2. It is observed that correct results can be got by this TDM workflow. The P-Bit will change its role in every 100 ns, and the simulation results stay in correct '1-1' or '-11' state. The shaded region in Fig. 17 indicates the wrong '11' or '-1-1' state. It shows that it can jump out from the wrong state very quickly. In addition, as shown in Fig. 18, the results for a XOR logic gate are plotted and 64% accuracy rate is achieved. Finally, the integer factorization for 35 is demonstrated in Fig. 19. 87% accurate rate for correct 5×7 and 7×5 results is clearly seen.

### V. Conclusions

In this work, a simple digital tunable TRNG based on STNO is proposed for the first time. The proposed tunable TRNG can be used as a P-Bit to comprise Ising computer. A novel incremental coupling rule is designed which coordinates well with our digital P-Bit device. Furthermore, a time division multiplexing scheme is proposed for Ising computer constructed by only one single P-Bit device. Our work sheds lights on a promising efficient hardware implementation of Ising computer.

### Acknowledgement

The authors wish to acknowledge the support from the National Key R&D Program of China (No. 2018YFB0407602), International mobility project under Grant B16001, National Key Technology Program of China under Grant 2017ZX01032101, National Postdoctoral Program for Innovation Talents under Grant BX20180028 and China Postdoctoral Science Foundation funded project under Grant 2018M641153.

## Tunable TRNG Based on STNO

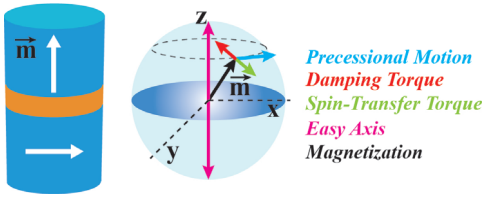

Fig. 1 The structure of STNO, and the illustration for the balance between anti-damping Spin Transfer Torque term and the Gilbert damping torque term.

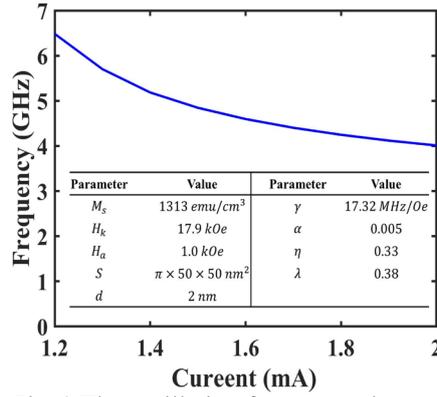

Fig. 2 The oscillation frequency changes with the injected current. Also shown in the table are the device parameters used in the simulation.

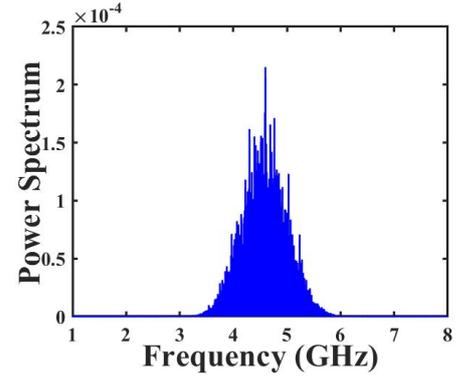

Fig. 3 The power spectrum of the FFT analysis for the STNO magnetic oscillation by room temperature. The frequency is widely spread in GHz.

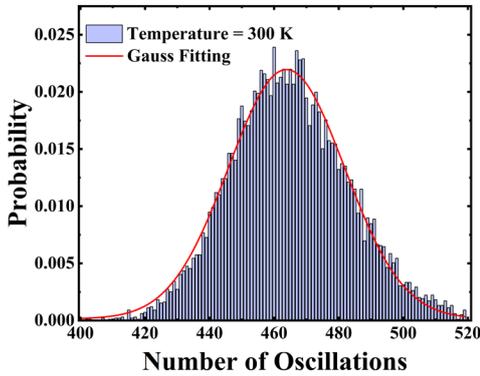

Fig. 4 The statistical analysis for the occurrence probability of the number of oscillations in every 100 ns. It is shown that a Gaussian function can fit it well.

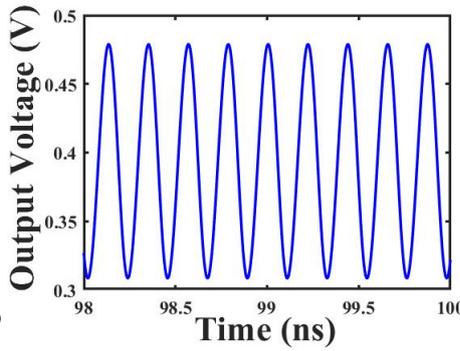

Fig. 5 The output voltage generated by the STNO with 1.6 mA injected current. A RF voltage output can be measured since the resistance varies with the magnetic rotation.

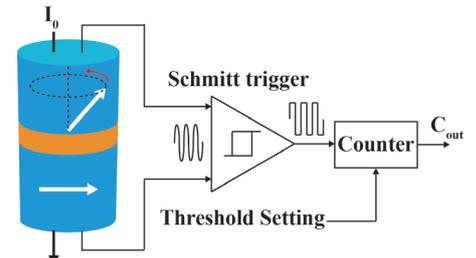

Fig. 6 The proposed digital TRNG based on STNO. A Schmitt trigger converts the RF output signal into square wave. A CMOS counter collects the number of oscillations and compares it with the predefined threshold. The threshold can be reset as needed.

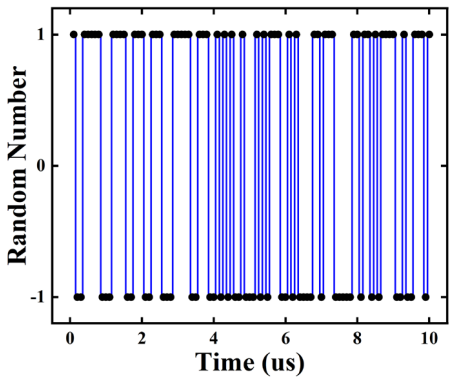

| TEST | P_value | | Pass/Fail |
|---|---|---|---|
| Approximate Entropy | 0.951292 | | Pass |
| Block Frequency | 0.112815 | | Pass |
| Cumulative Sums | 0.0602 (Forward) | 0.0602 (Reverse) | Pass |
| FFT | 0.218258 | | Pass |
| Rank | 0.69372 | | Pass |
| Frequency | 0.060341 | | Pass |
| Longest Run | 0.143519 | | Pass |
| Non-Overlapping Template | All sub-test : Pass | | |
| Overlapping Template | 1.000000 | | Pass |
| Runs | 0..071453 | | Pass |
| Serial | 0.44806 (P_value1) | 0.81712 (P_value2) | Pass / Pass |

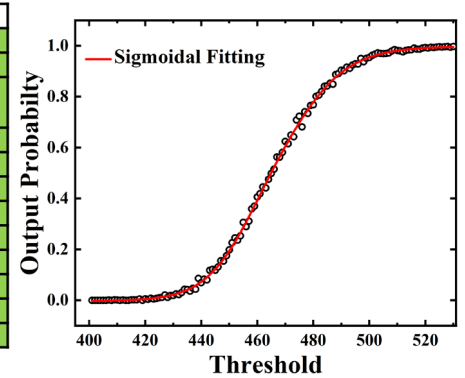

Fig. 7 The random number generated by the proposed digital TRNG. The 50/50 probability of '1' and '-1' is achieved when the predefined threshold is 466.

Fig. 8 The generated random number sequence can pass all of the 11 NIST tests.

Fig. 9 The output probability of the generated random number can be tuned by changing the threshold, and is fitted by the Sigmoidal function.

## Incremental Coupling Rule for Proposed P-Bit

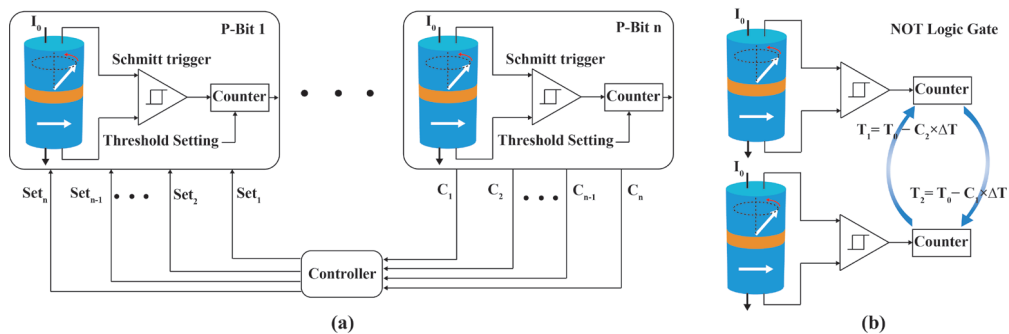

Fig. 10 (a) The illustration of the electrically connected P-Bits forming an asynchronous Ising computer. The controller is used to calculate the coupling strength. (b) The mutual coupling strength calculated from Eq. 1 for the two P-Bits comprising the NOT logic gate.

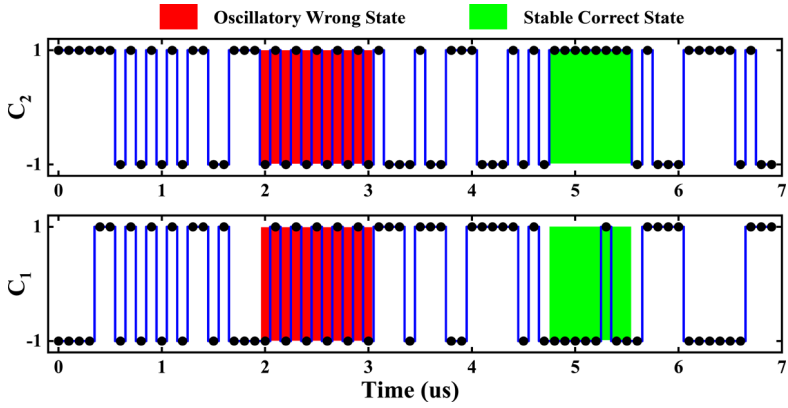
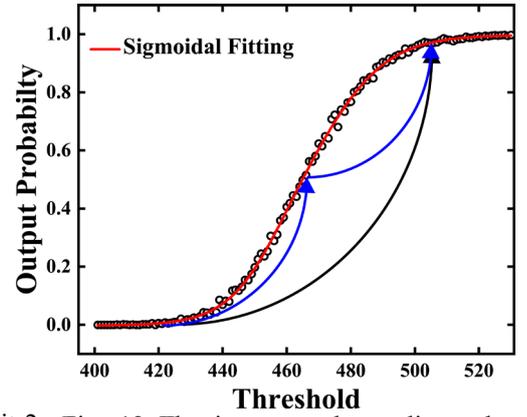

Fig. 11 The simulation results for the output of P-Bit 1 (Bottom) and P-Bit 2 (Top) of NOT logic gate with coupling rule Eq. 1. The accuracy rate is low since there exists oscillatory wrong state as shown in red region. In this wrong state, the result will oscillate between '11' and '-1-1'. While in the stable correct state shown in green region, the result is staying in '1-1' or '-11'.

Fig. 12 The incremental coupling rule: Instead of jumping between $T_{max}$ and $T_{min}$, the threshold of proposed P-Bit jumps from its previous value with a step of $\Delta T$.

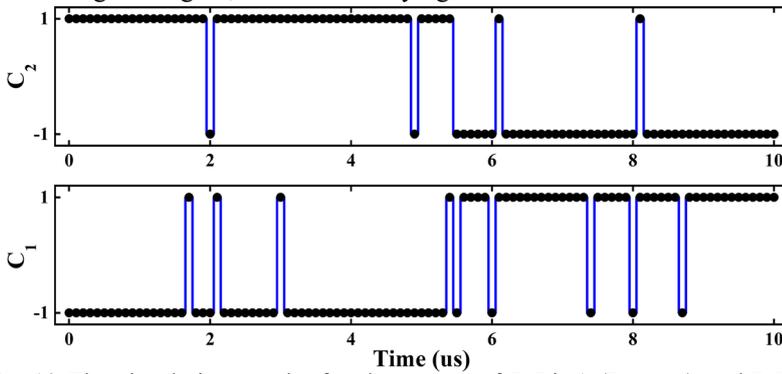
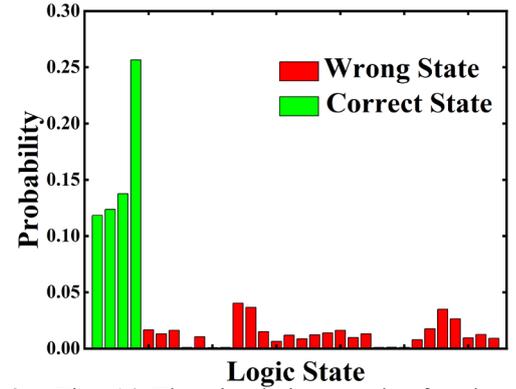

Fig. 13 The simulation results for the output of P-Bit 1 (Bottom) and P-Bit 2 (Top) of NOT logic gate with the incremental coupling rule Eq. 2. It is clearly seen that the oscillatory wrong state disappears. The accuracy rate increases to 74% when $T_{max} - T_{min} = 21$ and $\Delta T = 7$.

Fig. 14 The simulation results for the probability of each logic state occurrence for the XOR logic gate. Accuracy rate of 64% is achieved.

## Time Division Multiplexing Ising Computer

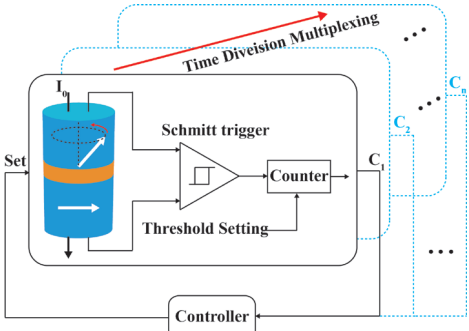
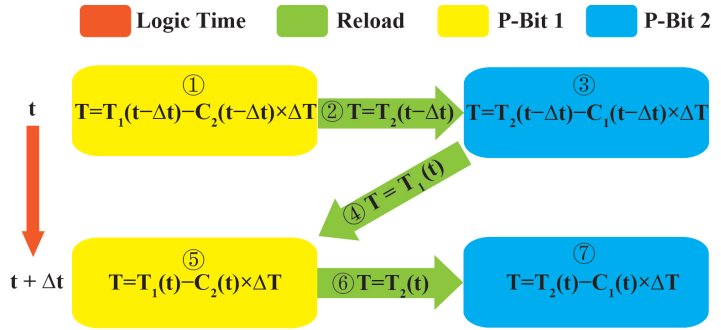

Fig. 15 The illustration for the time division multiplexing Ising computer with single P-Bit.

Fig. 16 The workflow for the time division multiplexing single P-Bits based on STNO to perform calculation.

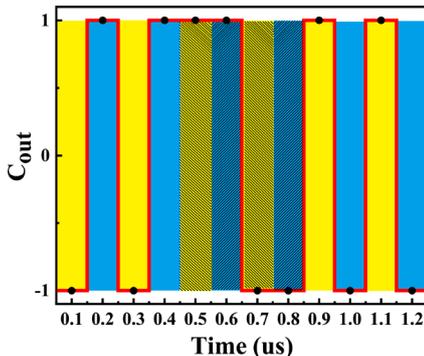
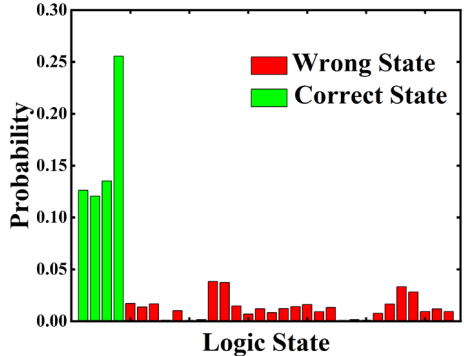
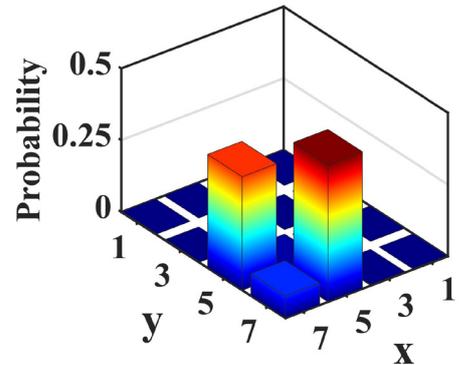

Fig. 17 The simulation results for TDM NOT logic gate. Yellow region presents P-Bit 1 and Blue region presents P-Bit 2.

Fig. 18 The simulation results for the probability of each logic state occurrence of TDM XOR logic gate.

Fig. 19 The simulation results for the probability of each combination of TDM integer factorization.